\documentclass[runningheads]{llncs}
\usepackage{graphicx} % Required for inserting images
\usepackage[T1]{fontenc}
\usepackage{mathtools}

\usepackage{array}
\usepackage{comment}
\usepackage{enumitem} % for customizing list environments
\usepackage{marvosym}
\title{Exploring Privacy Issues in Mission Critical Communication: Navigating 5G and Beyond Networks}

\titlerunning{Privacy in Mission Critical Communication}

\begin{document}

\author{Prajnamaya Dass\inst{1}\raisebox{0.5ex}{(\Letter)} \and % \orcidID{0009-0005-2589-6418}\and
Marcel Gräfenstein \inst{2}\and 
Stefan Köpsell\inst{1}}

\authorrunning{P. Dass et al.}
\institute{Barkhausen Institute, Dresden, Germany\\
\email{\{prajnamaya.dass, stefan.koepsell\}@barkhauseninstitut.org} 
\and
Technische Universität Dresden, Germany\\
\email{marcel\_daniel\_sven\_kevin.graefenstein@mailbox.tu-dresden.de}
}

\maketitle

\begin{abstract}
Mission critical communication (MCC) involves the exchange of information and data among emergency services, including the police, fire brigade, and other first responders, particularly during emergencies, disasters, or critical incidents. The widely-adopted TETRA (Terrestrial Trunked Radio)-based communication for mission critical services faces challenges including limited data capacity, coverage limitations, spectrum congestion, and security concerns. Therefore, as an alternative, mission critical communication over cellular networks (4G and 5G) has emerged. While cellular-based MCC enables features like real-time video streaming and high-speed data transmission, the involvement of network operators and application service providers in the MCC architecture raises privacy concerns for mission critical users and services. For instance, the disclosure of a policeman's location details to the network operator raises privacy concerns. To the best of our knowledge, no existing work considers the privacy issues in mission critical system with respect to 5G and upcoming technologies. Therefore, in this paper, we analyse the 3GPP standardised MCC architecture within the context of 5G core network concepts and assess the privacy implications for MC users, network entities, and MC servers. The privacy analysis adheres to the deployment strategies in the standard for MCC. Additionally, we explore emerging 6G technologies, such as off-network communications, joint communication and sensing, and non-3GPP communications, to identify privacy challenges in MCC architecture. Finally, we propose privacy controls to establish a next-generation privacy-preserving MCC architecture.

\end{abstract}

\section{Introduction}
Mission-critical services are the backbone of essential operations across various sectors, ensuring safety, security, and functionality in society. From police and fire brigade responses to healthcare delivery, transportation management, and industrial automation, these services play a pivotal role in safeguarding public welfare. Mission-critical communication (MCC) refers to the communication systems and technologies used to support and facilitate these essential functions and operations. Mission-critical communication systems are designed to provide reliable, secure, and resilient communication capabilities, particularly in situations where lives are at stake or where interruptions could have significant consequences.

The evolution of mission-critical communication began with land mobile radio (LMR) systems, which adhere to standards such as APCO Project 25 (P25) in the United States and TETRA (Terrestrial Trunked Radio) in Europe. These standards ensure interoperability between different LMR equipment and facilitate communication among various agencies and organisations. As technology advanced, digital mobile radio (DMR), which is based on the ETSI (European Telecommunications Standards Institute) standards, emerged as an alternative to traditional analog LMR, offering improved spectral efficiency, better voice quality, and support for features like encryption and data transmission. 

The architecture of both LMR and DMR systems was initially centred around voice-centric functionality due to its limited spectrum, coverage, and data capabilities. However, with the increasing demand for data-intensive applications, LTE gained traction in the mission critical communication domain due to its high data rates, low latency, and support for multimedia services. 3GPP (3rd Generation Partnership Project) in its technical specification (Rel-13) first defined standards and services for Mission-Critical Push-to-Talk (MCPTT) and in next release (Rel-14), added two additional services -- Mission-Critical Video (MCVideo) and Mission-Critical Data (MCData), within LTE that cater to MCC needs and paving the way for the eventual migration to 5G technology.

\subsection{5G-based mission critical communication (MCC) architecture}
\begin{figure} [h]
    \centering
    \includegraphics[width=1\linewidth]{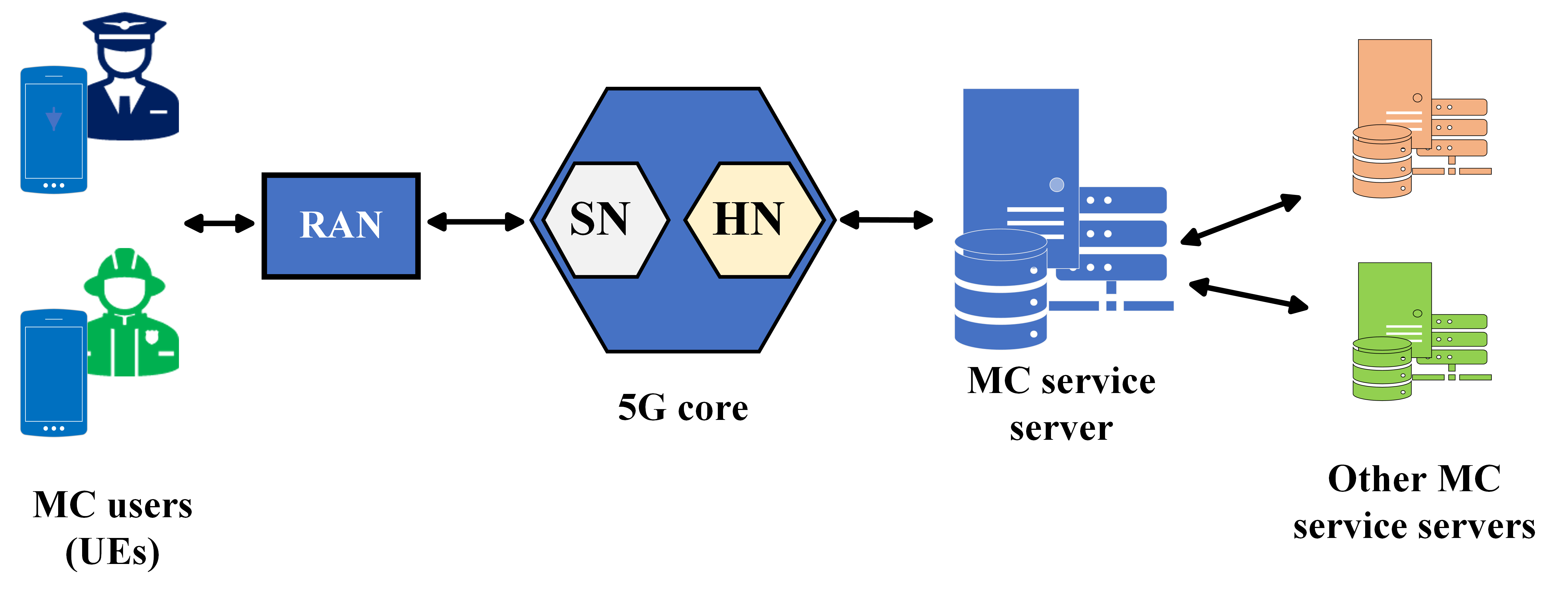}
    \caption{Overall mission critical communication architecture for 5G}
    \label{fig:mcc_arch}
\end{figure}

The architecture for mission critical communication, as depicted in figure \ref{fig:mcc_arch}, delineates a structured framework wherein mission critical users, who are existing subscribers of the network service provider (home network), utilise the network infrastructure to engage in communication with mission-critical (MC) servers \cite{3gpp_mcc_architecture}. Analogous to cellular users, both the home network (HN) and the serving network (SN) containing the radio access network (RAN) play pivotal roles in facilitating the initial connection for MC users to integrate with the network. Subsequently, each MC user initiates the establishment of a secure channel with the MC server, ensuring the integrity and confidentiality of communication. In instances where necessary, the MC service server retains the capability to establish connections between MC users and other MC service servers, thereby enabling seamless communication and collaboration within the mission critical system. In the core network, each network function is meticulously crafted with distinct purposes and functionalities tailored to efficiently manage and process various aspects of communication traffic and data. Each network function plays a crucial role in orchestrating the flow of information, managing network resources, and delivering high-quality services to end-users. For instance, the Authentication Server Function (AUSF) in 5G networks is responsible for authenticating and authorising user equipment (UE) when they attempt to access the network \cite{3gpp_5G_security}.

\section{Related Work}
In the literature, several works focused on the architectural concepts of mission critical communication and its security. The common functional architecture, procedures, and information flows to support mission-critical services over cellular networks are proposed by 3GPP \cite{3gpp_mcc_architecture}. The technical specification also specifies different possible deployment scenarios in which the functional MCC model can be applied. In \cite{RAN_MCC}, the challenges of using the 5G new radio interface for public safety MCC have been discussed. In \cite{feasibility_MCPTT}, feasibility issues of mission-critical push-to-talk (MCPTT) in 3GPP are discussed. In a similar work \cite{feasibility_MCC_4G}, the feasibility of MCPTT communications over 4G has been analyzed. The work \cite{MCPTT_5G} analyzes the impact of the evolution from 4G architectures toward 5G on MCPTT key performance indicators. Different transition possibilities of migration of mission-critical services from the land mobile radio-based systems to 4G \cite{LMR_LTE} and 5G have been discussed in \cite{next_gen_MC_networks}. The support of network slicing for MCC in 5G has been studied in \cite{MCC_networkslicing,MCC_networkslicing_5G}.

The security architecture and procedures to safeguard mission critical services have been specified by 3GPP in \cite{3gpp_mcc_architecture}. This specification outlines security mechanisms pertaining to on-network use, off-network use, roaming, and migration. Security threats to 5G interfaces have been analysed in \cite{5G_security_survey}, where standard security measures for these interfaces are discussed alongside categorised threats in their absence. The work in \cite{5G_privacy} discusses privacy threats in 5G stemming from newly introduced technologies like software-defined networking (SDN) and network function virtualisation (NFV).

\subsection{Motivation} The 3GPP technical specification for mission critical communication functional architecture and its security aspects mainly consider the underlying network (regardless of 4G or 5G technology) as the bridge to connect the MC users and the servers. However, the technologies introduced in 5G, such as network slicing and function virtualisation, introduce potential privacy concerns that are not considered in the current MCC architecture. While the 3GPP specifications for MCC address security concerns to some extent, privacy issues remain largely unexplored. While the existing security methods are necessary, they are not sufficient to address privacy issues. Additionally, the security and privacy solutions developed for 5G networks are not directly applicable to MCC architecture due to its distinct deployment scenarios.

With the integration of 5G networks into mission critical architecture, similar to typical 5G subscribers, various sensitive details of the MC users are disclosed to the network operator. For instance, the operator may obtain a policeman's current location, communication partners, and movement patterns over time, posing privacy risks. Moreover, application service providers within the MCC framework may exert control over user equipment and the MC client application, potentially extracting details like a policeman's identity and communication secret keys used with the MC server. Conversely, application and mission critical service providers may deduce sensitive information, such as the network ingress points and identifiers of network resources, posing privacy threats to the network as well.

Therefore, in this paper, we assess the privacy threats in MCC, while considering the 5G and the upcoming 6G technologies in the architecture. Considering different deployment scenarios, we show how an entity can extract or learn the personal identifiable information in the MC system. Additionally, we also suggest some privacy controls to deal with the privacy threats in the future mission critical systems.

% security is necessary but not sufficient; privacy aspects are still unaddressed
\section{Privacy Threats}

%In this section, we delve into the various threats in the MCC architecture that pose potential risks to any component within the MC system. These threats not only target the MC users but also encompass the network components and the MC server, rendering the entire system vulnerable to potential security breaches and privacy violations. During threat analysis, we primarily consider insider attackers, including the serving network (SN), home network (HN), MC users, and MC servers. Our investigation delves into how personally identifiable information of the MC system entities becomes exposed or acquired by other entities, thereby presenting privacy risks.

In this section, we delve into the various threats in the MCC architecture that pose potential risks to any component within the MC system. These threats not only target the MC users but also encompass the network components, the MC server, and their communications. Our threat analysis delves into how personally identifiable information of the entities in the MC system becomes exposed or acquired by other entities, thereby presenting privacy risks. 

\subsection{Privacy threats from the administering entities} 

\begin{figure} [h]
    \centering
    \includegraphics[width=1.2\linewidth]{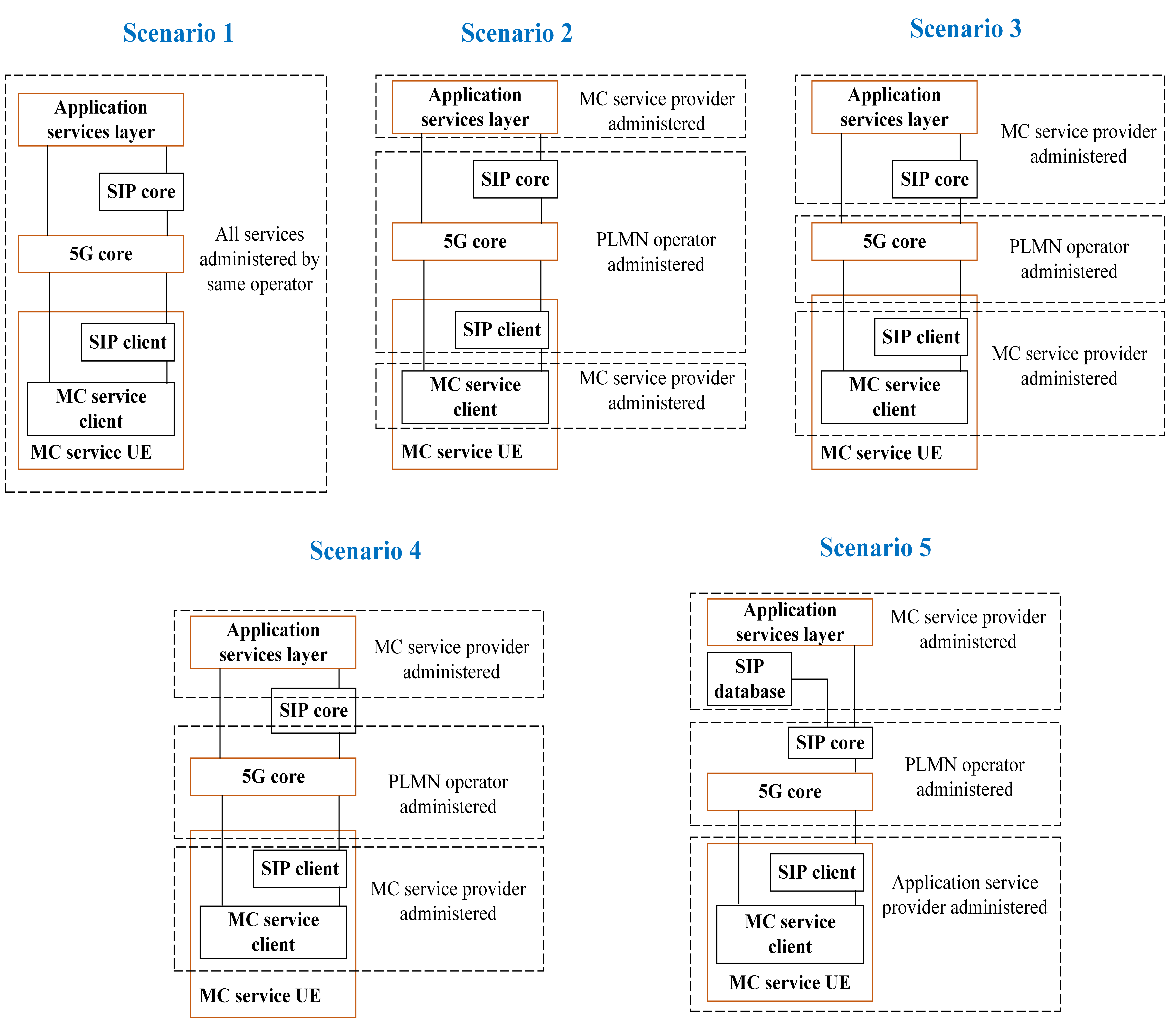}
    \caption{Different deployment scenarios for mission critical communication over 5G}
    \label{fig:deploment}
\end{figure}

Different deployment scenarios can arise based on the components or resources managed by either the mission-critical (MC) server or the public land mobile network (PLMN) operator \cite{3gpp_mcc_architecture}. Figure \ref{fig:deploment} illustrates five such deployment scenarios. 

In scenario 1, regardless of ownership, each resource is managed by the same operator. This scenario can unfold in three ways: i) the PLMN operator oversees network services, application services, and the MC service client, ii) the MC service server manages its own services and the underlying core network, iii) a third party, distinct from both the PLMN operator and the MC service server, manages all resources. However, in deployment scenarios 2, 3, and 4, some resources are managed by the MC service server while the remainder are overseen by the PLMN operator. In deployment scenario 5, the MC service user equipment is administered by the application service provider. It is important to note that a component may be owned by an entity different from the one administering it.

In mission critical communication, the majority of privacy threats can stem from administering entities. When an entity administers or controls a resource (whether software or hardware) or protocol, it often gains access to personal identifiable information about that resource or protocol. For example, in deployment scenario 5 (refer to figure \ref{fig:deploment}), if an MC service client, such as the client program in the user equipment, is administered by the application service provider, encryption keys and identities used by the MC client may be disclosed to the application service provider. Once in possession of identifiable and secret information, administering entities may launch spoofing and non-repudiation attacks. Likewise, in deployment scenarios (e.g., scenario 2 and 5 in figure \ref{fig:deploment}), where the session initiation protocol (SIP) and the communications over SIP are administered by the network operator \footnote{In this paper, the terms `PLMN operator' and `network operator' refer to the same entity.}, identifiable information such as the registered ID of the MC user in the SIP may be exposed to the network operator.

\subsection{Privacy threats during identity mapping and information sharing}
%The PLMN operator stores critical data, such as the location details of cellular subscribers, in the unified data repository (UDR) \cite{LCS}. Moreover, upon request, this information can be shared with the MC service server \cite{3gpp_mcc_architecture}. The process of location information request and response is shown in figure \ref{fig:identity_mapping}. For this information exchange to occur, the PLMN operator or the MC server must perform identity mapping to convert the MC user identity (MC ID) into the Subscription Permanent Identifier (SUPI). Unfortunately, this process exposes the MC ID to the PLMN operator, potentially enabling subsequent communications to be linked to the MC user. Consequently, this poses a privacy threat concerning identifiability and linkability. 

The PLMN operator, tasked with managing sensitive data such as the location details of cellular subscribers, maintains a centralised repository known as the unified data repository (UDR) \cite{LCS}. In instances where access to this data is necessary for mission-critical (MC) services, such as during emergencies or urgent communications, the UDR plays a pivotal role. Upon request, the PLMN operator facilitates the sharing of location information with the MC service server \cite{3gpp_mcc_architecture}, enabling swift and effective communication between MC users.

\begin{figure} [h]
    \centering
    \includegraphics[width=1\linewidth]{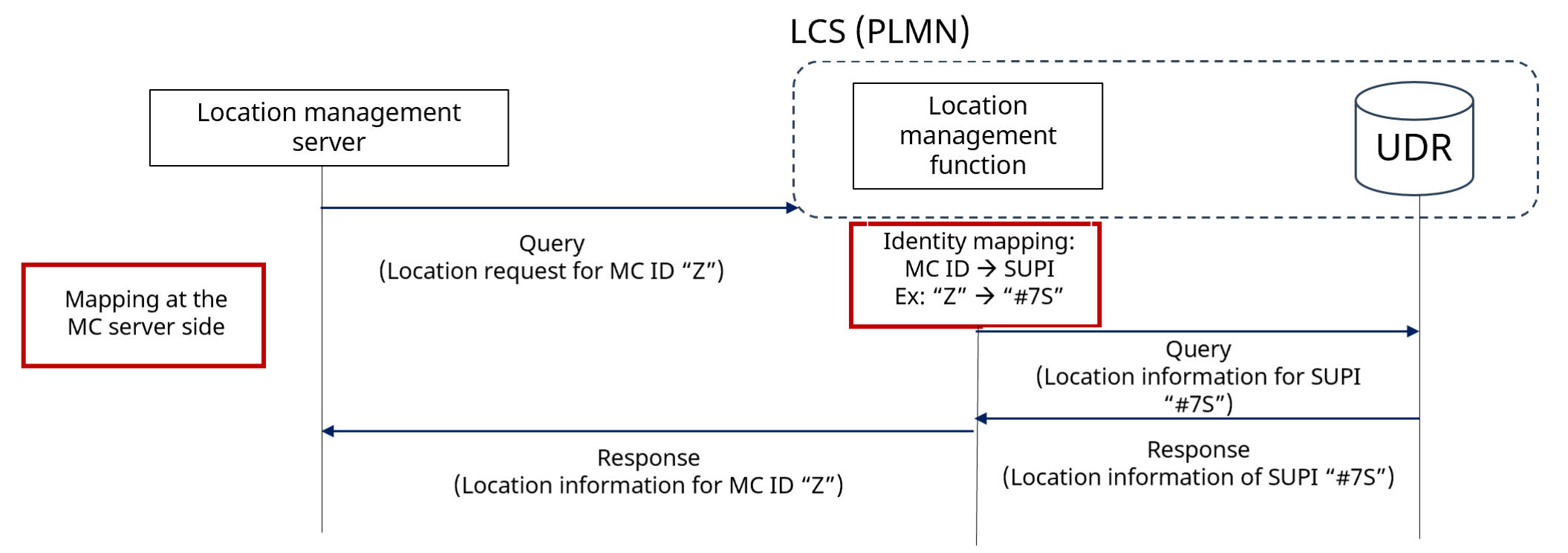}
    \caption{Privacy issues during information sharing between MC server and the PLMN operator}
    \label{fig:identity_mapping}
\end{figure}

The process of facilitating this information exchange involves a critical step known as identity mapping, depicted in figure \ref{fig:identity_mapping}. During this process, the PLMN operator or the MC server undertakes the conversion of the MC user's identity (MC ID) into the subscription permanent identifier (SUPI), a unique identifier associated with the subscriber's network subscription. While this conversion is essential for establishing the necessary communication channels, it inadvertently exposes the MC ID to the PLMN operator.

Unfortunately, this exposure of the MC ID to the PLMN operator raises significant privacy concerns, particularly regarding identifiability and linkability. By associating the MC ID with the SUPI, the PLMN operator gains insights into the identities of MC users, potentially enabling subsequent communications to be linked directly to individual users. This linkage not only compromises the anonymity and privacy of MC users but also creates a pathway for the aggregation of user-related data, raising concerns about the potential for unauthorised surveillance or tracking.

\subsection{Privacy risks arising from beyond trust domain interactions}
\begin{figure} [!ht]
    \centering
    \includegraphics[width=1\linewidth]{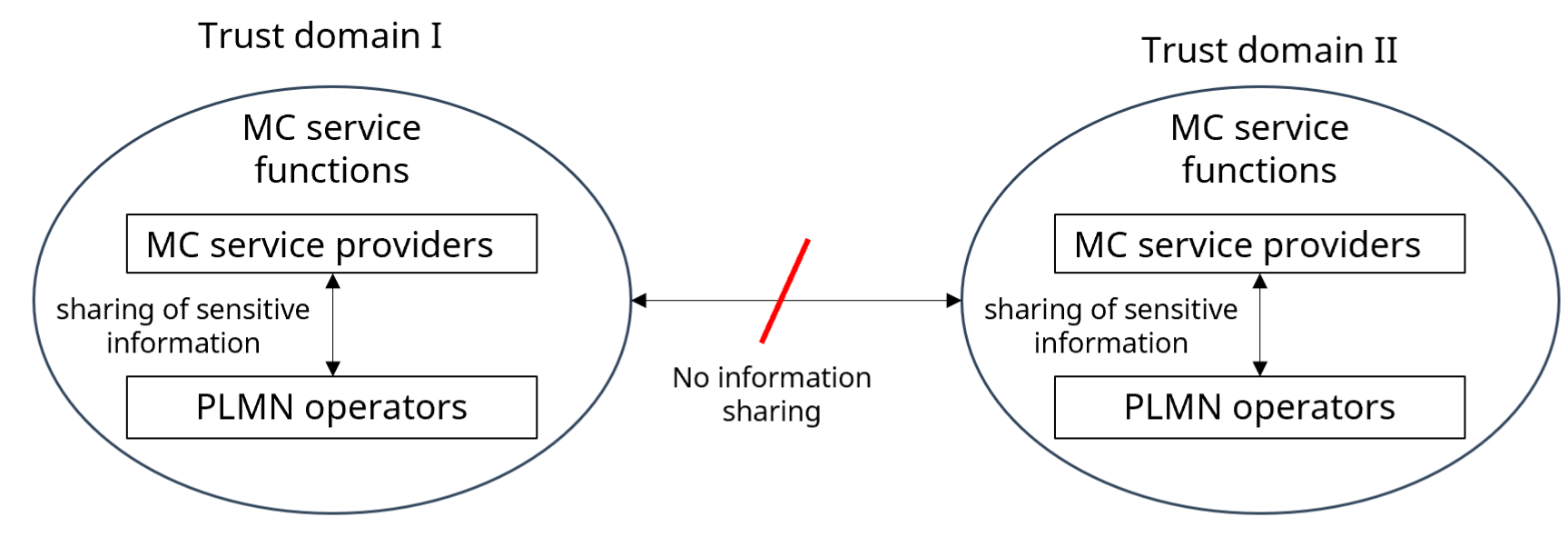}
    \caption{Restriction on information sharing inside and outside the trust domains}
    \label{fig:trust_domain}
\end{figure}

In an MC system, the trust domain encompasses one or more MC service functions managed by either the same or distinct service providers (such as the MC service provider or PLMN operator), who have agreed to exchange sensitive information, as shown in figure \ref{fig:trust_domain}. This implies that within a trust domain, a PLMN operator is restricted to sharing sensitive information exclusively with entities also within that same trust domain. However, given the limited number of PLMN operators in the market, establishing trust domains for information sharing presents a challenge. Additionally, if a single PLMN operator serves multiple MC service servers, delineating trust boundaries for the exchange of sensitive information becomes impractical. In such a scenario, an ill-intent PLMN operator common to two different trust domains can pose threats to linkability, identifiability, and non-repudiation.

\subsection{Privacy risks arising from priorities}
5G networks allow for prioritisation at different stages of the communication establishment. Thereby MC users and the related services have higher priority than ``normal'' users to guarantee availability of the network and its service to MC users even in case the network is overloaded. However, from privacy perspective, this prioritisation can be used to identify MC users.

At the user plane, i.e., related to the data transport, a so-called quality of service flow ID (QFI) is assigned to every data session. The QFI value, known to entities such as the RAN and SN, can be used to identify mission critical communication. For instance, QFI value 65, which is assigned to ``Mission Critical user plane Push-To-Talk voice'' (please refer to Table 5.7.4.1-1 in~\cite{3gpp_architecture}), gives identifiable information of an MC user.    

The concept of Unified Access Control deals with an overloading situation at the RAN. Therefore, every establishment of a radio session between an UE and the RAN contains an Access Category and one or more Access Identities (AI). The AI value for mission critical services is 2. It allows MC users to get access to the RAN even if the RAN is currently overloaded. Therefore, ``normal'' users will be disconnected to allow service to MC users.

Finally, during the connection establishment, a value referring to the cause of the establishment (EstablishmentCause in RRCSetupRequest) is specified in the request \cite{5G_NR_3GPP}. Thereby, special values exist to signal that specifies the establishment of the connection is related to mission critical services. By analysing these values (QFI, AI, and EstablishmentCause), the network operators can get rich contextual information about mission critical communication. This includes insights into the number of MC users present under a specific base station, their communication patterns, and the types of services they are utilising, which can lead to linkability and identifiability threats.

\subsection{Threats from the session initiation protocol (SIP)}
Privacy threats can arise from both SIP clients that reside in the MC service UE and SIP core infrastructure in the mission critical communication systems. The SIP client is the application in the UE responsible for initiating or receiving SIP sessions, while the SIP core, also referred to as the SIP infrastructure (includes proxy servers, SIP server, application servers, etc.) is the network element that routes SIP messages within a SIP network. SIP clients, responsible for initiating and managing communication sessions, may inadvertently disclose sensitive user information, including identity details and communication patterns of the MC user. Such leaks could be exploited by adversaries for profiling or targeted attacks, compromising user privacy. Moreover, insecure SIP clients are vulnerable to eavesdropping and identity spoofing, enabling unauthorised parties to intercept communications or impersonate legitimate users, further undermining the integrity and security of the system. Within the literature, various solutions have been proposed to anonymise header information in SIP \cite{RFC_SIP_privacy}. Nonetheless, identifiable details regarding the MC user are still revealed in instances where the SIP client is managed either by the PLMN operator (in deployment scenarios 1 and 2) or by another service provider (in deployment scenario 5).

Similarly, vulnerabilities in the SIP core infrastructure introduce privacy risks, potentially leading to unauthorised access to user data or call records. Adversaries could exploit these weaknesses to compromise user privacy, conduct traffic analysis attacks, or disrupt services through denial-of-service (DoS) attacks. When the SIP core is administered by entities other than the MC service server (in deployment scenarios 2, 4, and 5), identifiable information of the MC service server about the server as well the MC users are revealed to the administering entity.

\subsection{Privacy issues during secure tunnel establishment}
%Transport Layer Security (TLS) facilitates a secure connection between the MC client and server. However, prior to TLS initiation, DNS (Domain Name System) mapping occurs through recursive solvers and name servers. When a subscriber initiates a connection request, the recursive solver, managed by the PLMN operator, translates the destination address into an IP address. If the destination corresponds to an MC server, the PLMN operator deduces that the subscriber likely belongs to the MC service. This poses a threat to linkability and identifiability, potentially exposing sensitive information about the MC user. If the PLMN operator controls the name servers, similar risks arise during DNS mapping. Although TLS secures connections from third-party interference, using proxy servers can anonymise the source address but not the destination address. Additionally, the PLMN operator may glean significant information, such as which requests target specific MC servers, distinguishing between MC and regular 5G users via SUPI, and identifying the number and locations of MC users. Furthermore, contextual details like communication timing and group activities may also be accessible to the PLMN operator.

Transport layer security (TLS) plays a crucial role in establishing a secure connection between the MC (mission critical) client and server, ensuring that sensitive data exchanged between them remains encrypted and protected from eavesdropping or tampering. However, before TLS comes into play, the process of DNS mapping unfolds, orchestrated by recursive solvers and name servers.

When a subscriber initiates a connection request, the recursive solver, under the administration of the PLMN operator, undertakes the task of translating the destination address into an IP address. This translation step is pivotal for routing the communication to its intended destination. However, it also presents a potential privacy concern. If the destination address corresponds to an MC server, the PLMN operator can infer that the subscriber is likely a participant in the MC service. Note that even without the DNS-based hostname resolution, the nature of the communication target might be derived just from the destination IP address. This inference poses a threat to the linkability and identifiability of the user, as it opens avenues for exposing sensitive information associated with the MC user's activities and affiliations.

Moreover, if the PLMN operator wields control over the name servers involved in the DNS mapping process, similar risks arise. This control could potentially exacerbate privacy concerns, as the operator gains more insights into the communication patterns and preferences of its subscribers. While TLS does provide a layer of protection against third-party interference by encrypting the communication channel, it's worth noting that it primarily safeguards the content of the communication rather than its metadata. In scenarios where proxy servers are employed, while they may anonymise the source address, the destination address remains visible, thus posing a challenge to preserving the anonymity of users.

Furthermore, beyond the encryption provided by TLS, the PLMN operator may still gather significant insights into the communication ecosystem. This includes discerning which requests are directed towards specific MC servers, distinguishing between MC users and regular 5G users via identifiers like SUPI (Subscription Permanent Identifier), and even tracking the distribution and concentration of MC users across different geographical locations.

In addition to these aspects, contextual details such as the timing of communication and group activities may also be within the purview of the PLMN operator's surveillance capabilities. This wealth of information not only raises concerns regarding user privacy but also underscores the importance of implementing robust safeguards and regulations to ensure the responsible handling of sensitive data in mission-critical communication environments.

\subsection{Threats from network slicing}
\begin{figure}[h]
    \centering
    \includegraphics[width=1.0\linewidth]{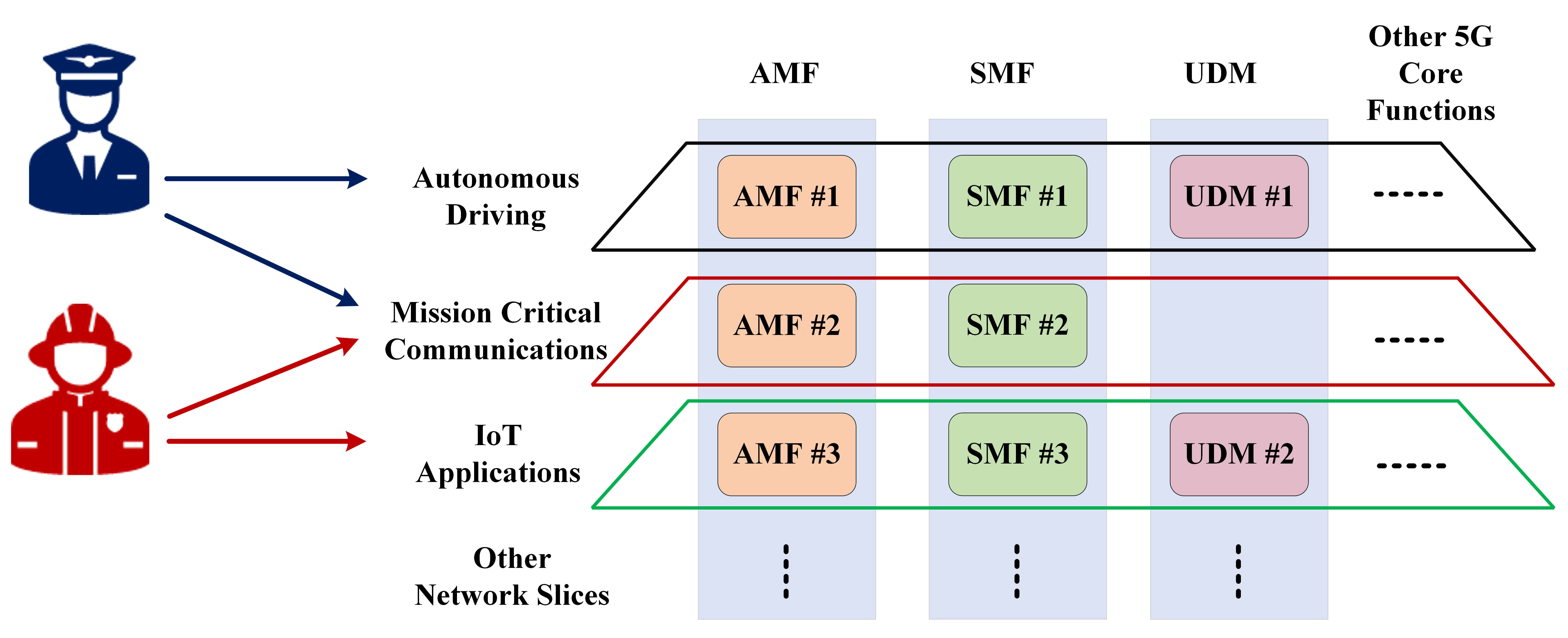}
    \caption{Privacy threats due to network slicing}
    \label{fig:network-slicing}
\end{figure}

The network slicing is an architectural concept that divides a single physical network infrastructure into multiple virtual networks, each tailored to specific use cases, applications, or customer requirements \cite{network_slicing}. As shown in figure \ref{fig:network-slicing}, a network slice may be allocated to mission critical communications. Network slicing, while offering significant benefits in terms of customisation, flexibility, and resource allocation in 5G and beyond, also introduces several security and privacy threats \cite{5G_threat_landscape}.

An MC (mission critical) user, subscribed to a network slice dedicated to critical services, engages in communication that involves sensitive information, such as location data and MC organisation information. Data intended for the MCC slice might inadvertently leak into other slices, exposing personally identifiable information about MC users, network entities, or MC server entities to unauthorised parties. Network operators or malicious actors may conduct traffic analysis across different slices to identify patterns or behaviours associated with MC users. By analysing traffic patterns or metadata, adversaries may gain insights into the behaviour or activities of MC users even when they are present in other slices. This could compromise user privacy and security, as it reveals information about their interactions, preferences, or usage patterns. Unauthorised access to slice-related information, such as the participation of specific network functions (NFs) common in MCC and other slices, could lead to privacy breaches or security incidents.

An MC user is also part of another slice, perhaps a slice designated for IoT devices or enterprise services. The user's identity is tied to a subscription permanent identifier (SUPI), which is shared across slices. The network operators could exploit SUPIs or other identifiers to correlate the identities of users across different slices. This linking of identities could lead to privacy violations, as it enables adversaries to aggregate and analyse PII from multiple sources, potentially revealing sensitive information or behavioural patterns.

\subsection{Threats from the 5G authentication protocols} 

The authentication protocol in 5G networks, known as 5G AKA (Authentication and Key Agreement) \cite{3gpp_5G_security}, is susceptible to several privacy threats, such as linkability and traceability due to its design and implementation \cite{5G_AKA_privacy,5G_AKA_privacy2}. The subscription concealed identifier (SUCI) is used as an essential element in the authentication process instead of the subscription permanent identifier (SUPI) utilised in previous generations. However, the use of SUCI does not entirely eliminate privacy threats in the authentication protocol; rather, it presents its own set of challenges. Further, due to the involvement of sensitive services from the MC server, the 5G AKA protocol introduces some more threats.

\begin{figure}
    \centering
    \includegraphics[width=0.9\linewidth]{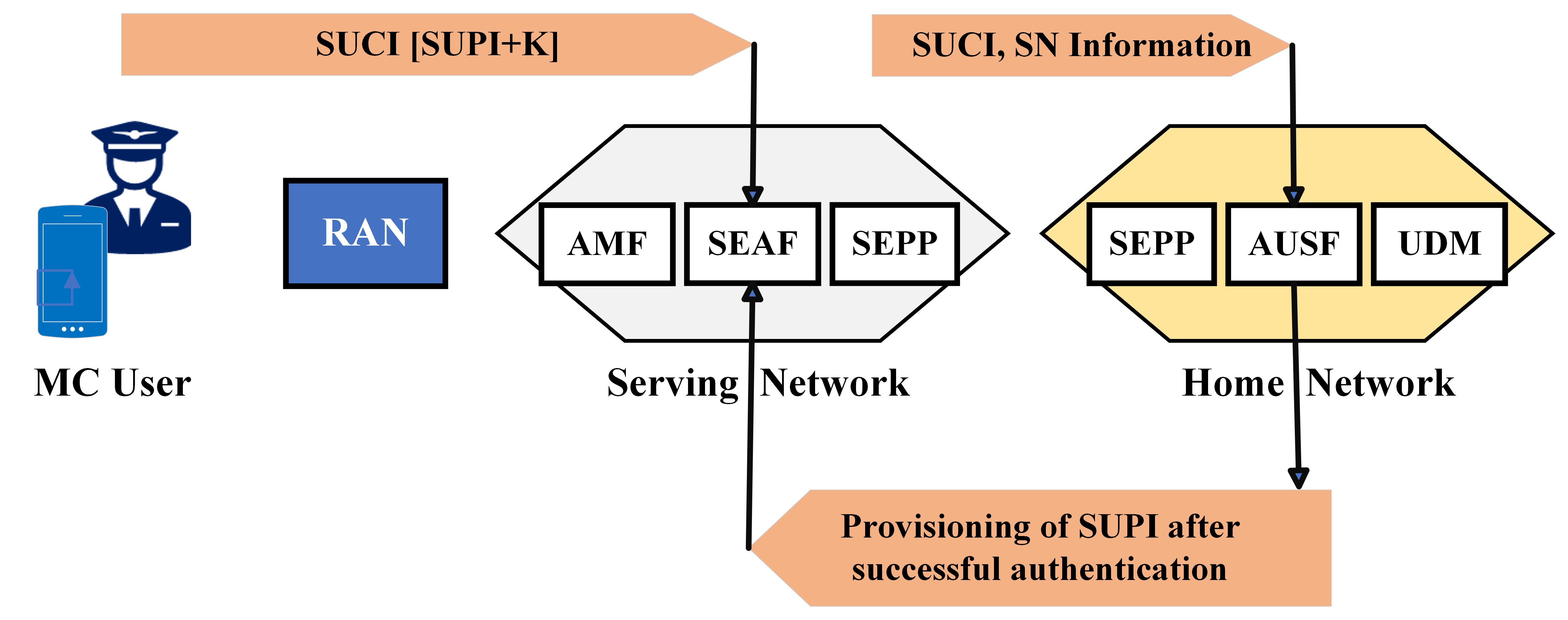}
    \caption{Authentication process during the 5G registration procedure}
    \label{fig:SUPI_disclosure}
\end{figure}

After successful authentication in 5G AKA, the SUPI is exchanged with the serving network \cite{5G_AKA_privacy2}, as shown in figure \ref{fig:SUPI_disclosure}. This exchange is a crucial legal step in establishing the user's connection to the network and enabling access to services. However, the sharing of SUPI introduces privacy considerations. The exchange of SUPI between the MC user and the serving network (SN) exposes the user's permanent identifier to entities of the SN. While necessary for network operations, this exposure can potentially be exploited by the SN and malicious actors to track the activities of the MC users. SUPI, being a permanent identifier tied to the user's subscription, can be used to uniquely identify and track users across different communication sessions or contexts. This raises concerns about user privacy and anonymity.

\subsection{Privacy challenges in the upcoming 6G technologies}
The introduction of new technological features like off-network communication (using proximity-based services or ProSe) \cite{3gpp_mcc_architecture} and joint communication and sensing (JCAS) \cite{JCAS_usecase} in 6G networks can potentially pose privacy threats. Here's how each of these features could impact privacy:

\begin{itemize}
    \item \textbf{Off-network communication}: ProSe enables devices to communicate directly with each other when they do not have connectivity with the RAN or the network infrastructure. Figure \ref{fig:off_network} represents the on- and off-network mission critical communication. Off-network communication is a part of the future MC services \cite{3gpp_mcc_architecture,3gpp_mcc_security}. While this offers convenience and efficiency, Off-network communication without active support from MC server, is more prone to security and privacy threats. As devices exchange information directly, their proximity to each other can be inferred, potentially revealing sensitive information about MC users' locations and movements. ProSe communications may expose device identifiers or MC identities, compromising user anonymity and privacy. Adversaries may exploit this information to track users' activities or identify individuals participating in ProSe interactions. Due to the absence of a link with the core network and the MC server, the direct communication between the MC users may bypass traditional security measures implemented by network operators, increasing the risk of data leakage. In addition to network operators, other MC users serving as relays or being available to nearby communicating MC users may also have the ability to infer sensitive information.
    
    \begin{figure}[!ht]
    \centering
    \includegraphics[width=1\linewidth]{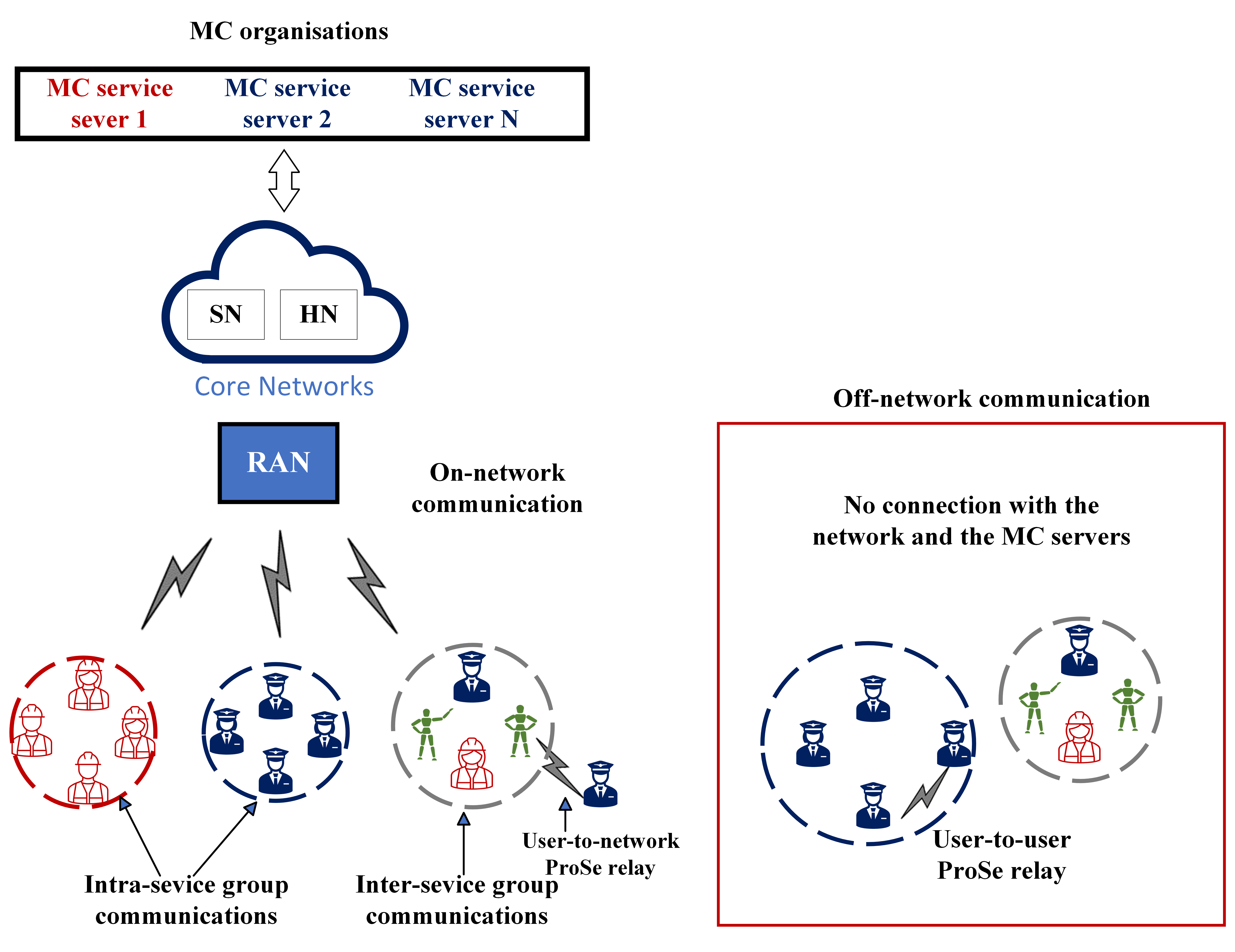}
    \caption{On-network and off-network mission critical communications}
    \label{fig:off_network}
    \end{figure}

    \item \textbf{Joint communication and sensing (JCAS)}: The utilisation of a single radio signal for both communication and sensing functions is commonly referred to as integrated or joint communication and sensing \cite{JCAS_def}. Consequently, it unlocks fresh possibilities for applications necessitating both communication and sensing functionalities, particularly for platforms previously incapable of accommodating both simultaneously. Given the nature of sensing data involved, which often includes sensitive personally identifiable information related to humans and objects, JCAS-based applications are particularly susceptible to privacy attacks, such as location tracking, identity disclosure, profiling, and misuse of sensor data. With the incorporation of sensing capabilities, PLMN operators can pinpoint the location of MC users accurately, provided they can differentiate them from other users. Additionally, the aggregation of sensing information from multiple MC users may enable the creation of detailed profiles of individuals' behaviors, preferences, and activities. This data aggregation raises concerns about MC user profiling, algorithmic discrimination of MC services, and potential misuse of personal information by third parties.
    
    \item \textbf{Non-3GPP access}: As the scope of IoT applications and use cases continues to broaden, forthcoming generations are expected to experience a notable surge in the volume of devices linking to the Internet. In contrast to mobile phones, which are classified as 3GPP devices with cellular connectivity, non-3GPP user equipment, such as IoT devices, sensors, and wearables, utilise Wi-Fi or alternative wireless technologies for Internet or local network connectivity \cite{non-3GPP,non-3GPP_access}. In mission-critical services, there is a provision for non-3GPP devices to connect to the MC server through non-3GPP and 3GPP access \cite{3gpp_mcc_architecture}. As shown in Figure \ref{fig:non-3GPP-access}, the non-3GPP devices, which cannot host MC clients, use the MC gateway (MC UE) to establish communication with the MC server. One non-3GPP device can have multiple MC clients (e.g., one for MCPTT and another for MCVideo), which can use different UEs as gateways to connect to multiple MC servers.
    
    \begin{figure} [!ht]
    \centering
    \includegraphics[width=1\linewidth]{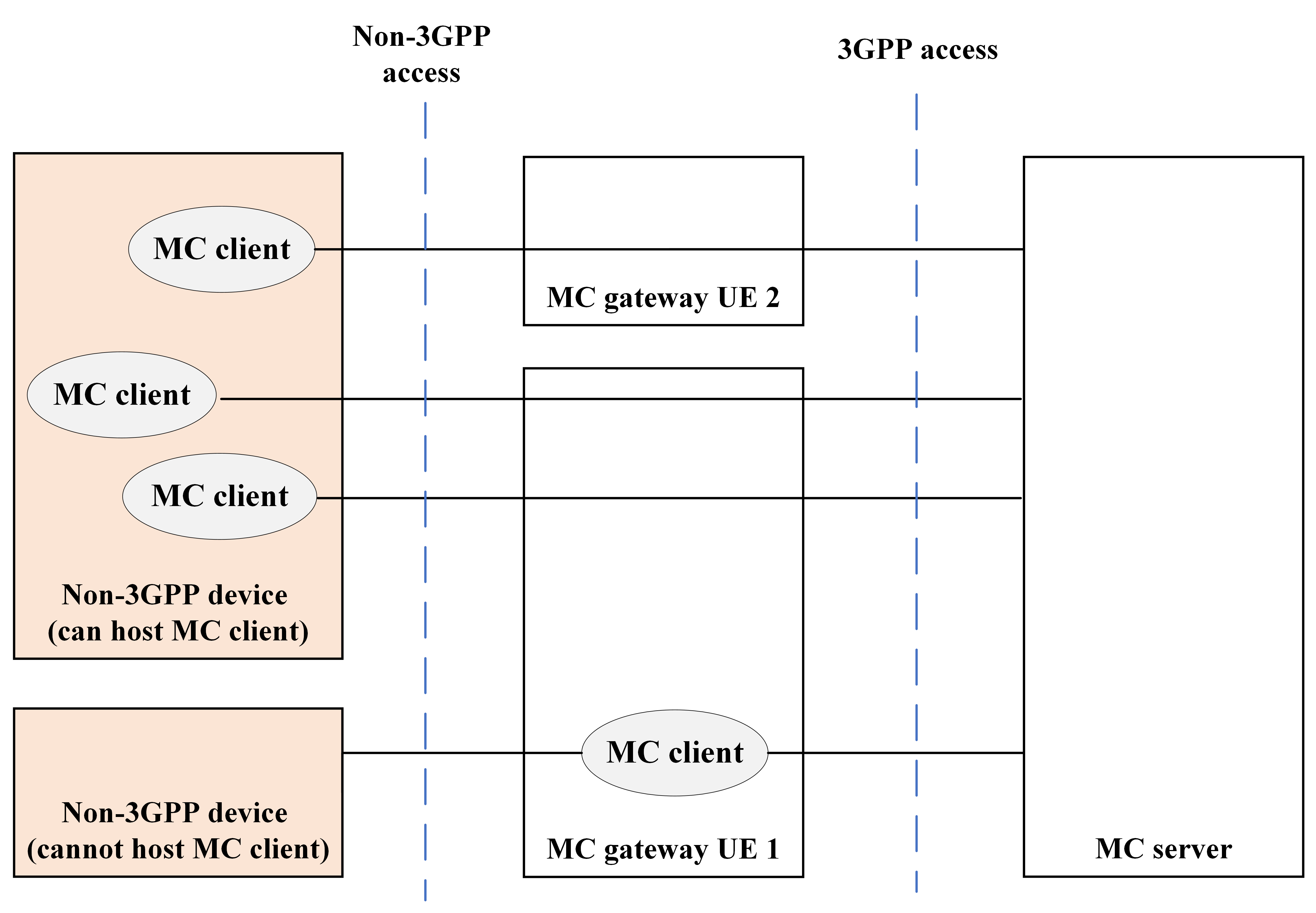}
    \caption{Communication between non-3GPP MCC devices and the MC server}
    \label{fig:non-3GPP-access}
    \end{figure}

    The connection of non-3GPP devices to the MC UEs through non-3GPP access, presents a range of privacy threats stemming from various factors. These devices often collect and transmit sensitive mission critical data over unsecured channels, making them susceptible to interception or unauthorised access. In addition, they may rely on communication protocols lacking robust security mechanisms, leaving the MC user communications vulnerable to eavesdropping or man-in-the-middle attacks. Non-3GPP devices often rely on third-party services or cloud platforms for data storage, processing, and analysis. If used for mission critical services, these services may not adequately protect MC user's data or may have lax security practices, increasing the risk of data breaches or unauthorised access. Further, the non-3GPP service provider may acquire certain information about the MC UE. In scenarios, where the PLMN operator manages certain components of the MC server (deployment scenario 5 in figure \ref{fig:deploment}), information about the non-3GPP devices could be revealed to the PLMN.
     
\end{itemize}

\section{Suggestive Privacy Controls}

In the 3GPP standard for mission critical communication (MCC) security architecture \cite{3gpp_mcc_security}, numerous controls have been established to uphold robust security measures. However, it's crucial to note that these proposed security aspects primarily align with the 4G network architecture and may not fully incorporate the advancements and innovations introduced in the 5G landscape. Although many privacy issues are challenging to solve, in this section, we discuss possible solutions to counter some of the privacy issues discussed in this paper.

\begin{itemize}

\item \textbf{Privacy-preserving information retrieval}: Several methods can be employed to retrieve information, such as the location history of MC user from the core network, without revealing the real identity designated to the user by the MC server. It should be noted that the privacy-preserving information retrieval must ensure authentication and authorisation checks. Private information retrieval (PIR) schemes \cite{PIR_survey,PIR_singledb} ensure confidentiality and privacy during information retrieval by concealing the user's query from the database server. Therefore, PIR protocols could be used to retrieve information from the unified data repository (UDR) in the core network without revealing which MC user information records are being accessed. Authentication methods, such as blind signatures \cite{blind_signature}, anonymous authentication \cite{anonyous_authentication}, and zero-knowledge proofs \cite{zero-knowledge-proof}, could be used to verify the authenticity of the MC server requesting information from the UDR. Furthermore, the use of anonymous communication networks that route user traffic through a series of encrypted relays, such as Tor (The onion router) \cite{Tor}, can help preserve the anonymity of the origin of the location information request (MC server).

\item \textbf{Privacy in inter-trust domain communications}: Ensuring privacy within trust domains, especially when the same PLMN operator spans multiple trust domains, requires a multifaceted approach. At the very first step, strict compliance with relevant privacy regulations and standards, such as the General Data Protection Regulation (GDPR) is required to adhere to legal requirements for data sharing, processing, and protection across trust domains. Further, it is also needed to maintain transparency with data subjects about how the mission critical data is shared and processed across trust domains. The best way is to limit the amount of personally identifiable information of MC services shared to the PLMN operators, which is necessary for the intended purpose. It is suggested to anonymise or pseudonymise the PIIs used in MC services to remove or obfuscate identifiers that can link data to the entities involved in MCC. One-time tokens are designed with a brief lifespan, intended for single use or a limited duration \cite{one-timeprogram}. Once employed within a trust domain, the token immediately becomes invalid, thereby diminishing the potential for unauthorised access or exploitation in alternate trust domains. These controls will also help in achieving privacy against third-party service providers.

\item \textbf{Privacy controls for MCC network slice}: Ensuring privacy of MCC in network slices, especially when the core network functions are shared among multiple slices, requires a comprehensive approach that combines technical measures and governance policies. Strict data segregation mechanisms prevent cross-contamination between slices, if the MCC network slice has its own dedicated data storage and processing resources. Implementing granular access controls can help to regulate access to data and resources within the MCC network slice. Use of role-based access control (RBAC) \cite{RBAC} or attribute-based access control (ABAC) will help to enforce least privilege principles and restrict access to authorised users only \cite{ABAC}.

\item \textbf{Privacy controls for upcoming 6G technologies}: During off-network communication and sensing activities in JCAS, techniques such as masking and randomisation can obscure identifiable information of the MC users participating in mission critical communications. Achieving unlinkability and unobservability for ProSe-based mission critical communications can be attained through covert operations, noise injection, anonymisation, data fragmentation, decoy traffic, and differential privacy techniques. Employed individually or in combination, these methods aim to obscure and obfuscate intra and inter-service group communications.

\end{itemize}

\section{Conclusions and Future Work}
Privacy in mission critical communication is essential, especially considering the involvement of sensitive information concerning public safety, national security, and critical infrastructure. The role of network operators and application service providers in enabling such communication is notably significant, underscoring the necessity for rigorous privacy safeguards. In this paper, we examined the privacy aspects of mission critical communication in context to 5G networks, discussing the architecture and analysing weaknesses posing privacy threats to MC users, network entities, and MC servers. Moreover, we extended our analysis to anticipate potential privacy threats stemming from the advent of 6G technologies, such as off-network communication, joint communication and sensing, and the integration of non-3GPP device communication. In response to these evolving challenges, we suggested a suite of privacy controls tailored to mitigate current and future threats, with a forward-looking approach towards addressing next-generation privacy concerns. 

The current mission critical communication architecture specifications do not consider the underlying 5G technology and protocols. Looking ahead, we plan to delve deeper into protocol and interface-level threat analysis in the mission critical communication architecture, with a commitment to identifying and offering highly specific and detailed privacy-preserving solutions to safeguard this critical domain.

\section*{Acknowledgement}

This work has been partly funded by the German Federal Office for Information Security (BSI) project 6G-ReS (grant no. 01MO23013D). Additionally, the authors from Barkhausen Institute are also financed based on the budget passed by the Saxonian State Parliament in Germany. 

\begin{comment}
\begin{table} [!ht]
  \centering 
   \renewcommand{\arraystretch}{1.7} % Increase the spacing between rows
    \caption{Summary of threats in mission critical communication over 5G}
    %\label{tab:threat_table}
    \scalebox{0.8}{%
    \begin{tabular}{|m{6cm}|m{5cm}|m{4cm}|} % adjust column types and widths as needed
        \hline
        \textbf{Weakness or reason} & \textbf{Threat type} & \textbf{Deployment scenario} \\
        \hline
        Disclosure of SUPI of the MC user to the serving network & 
        Linkability, Identifiability, Detectability, Non-repudiaion & 
        All deployment scenaios \\  
        \hline 
        Inadequate security mechanism for mapping MC user identity to SUPI & 
        Linkability, Identifiability, Detectability & 
        All deployment scenaios \\  
        \hline 
        Disclosure of session initiation protocol (SIP) identity to the PLMN operator or third parties &
        Linkability, Identifiability, Detectability &
        2, 4, 5 \\
        \hline 
        Disclosure of PIIs used in one network slice with other network slices &
        Linkability, identifiability, Detectability &
        All deployment scenarios \\
        \hline
    \end{tabular}}
\end{table}
\end{comment}
\bibliographystyle{splncs04}
\bibliography{references}

\end{document}